\documentclass[a4paper,12pt,titlepage]{article}
\usepackage[utf8]{inputenc}
\usepackage{amsmath}
\usepackage{amsfonts}
\usepackage{amssymb}
\usepackage{float}
\usepackage{graphicx}
\usepackage{setspace}
\usepackage{adjustbox}
\usepackage{xcolor}
\usepackage{csquotes}
\usepackage{listings}
\usepackage{lscape} 
\usepackage{threeparttable}
\usepackage[backend=bibtex, style=authoryear, maxcitenames=3, maxbibnames=6, minbibnames=6, giveninits=true, uniquename=init]{biblatex}
\RequirePackage[rmargin=1in,lmargin=1in, lines=25]{geometry}
\usepackage{authblk}
\newcommand{\ubar}[1]{\mkern3mu\underline{\mkern-3mu #1\mkern-3mu}\mkern3mu}

\usepackage{subfiles} % Best loaded last in the preamble

\title{}
\date{}

\begin{document}
\thispagestyle{plain}
\begin{center}
    {\Large
    \textbf{Long-term memory effects of an incremental blood pressure intervention in a mortal cohort}}
        
\vspace{0.1cm}
\textsc{Josefsson, Maria$^{1,2,*}$, Karalija, Nina $^{2,3}$, \& Daniels, Michael J.$^4$}
\end{center}

\begin{center}{\footnotesize
$^1$Department of Statistics, USBE, Ume{\aa} University, Sweden.

\vspace{-0.05cm}
$^2$Ume{\aa} Center for Functional Brain Imaging, Ume{\aa} University, Sweden.

\vspace{-0.05cm}
$^3$Department of Integrative Medical Biology, Ume{\aa} University, Sweden.

\vspace{-0.05cm}
$^4$Department of Statistics, University of Florida, USA.

\vspace{-0.1cm}
$^*$Corresponding author, E-mail: maria.josefsson@umu.se}
\end{center}

%%%%%\begin{center}
\textbf{Summary}
In the present study we investigate overall population effects on episodic memory of an intervention over 15 years that reduces systolic blood pressure in individuals with hypertension. A limitation with previous research on the potential risk reduction of such interventions is that they do not properly account for the reduction of mortality rates. Hence, one can only speculate whether the effect is due to changes in memory or changes in mortality. Therefore, we extend previous research by providing both an etiological and a prognostic effect estimate. To do this, we propose a Bayesian semi-parametric estimation approach for an incremental threshold intervention, using the extended G-formula. Additionally, we introduce a novel sparsity-inducing Dirichlet hyperprior for longitudinal data, that exploits the longitudinal structure of the data. We demonstrate the usefulness of our approach in simulations, and compare its performance to other Bayesian decision tree ensemble approaches. In our analysis of the data from the Betula cohort, we found no significant prognostic or etiological effects across all ages. This suggests that systolic blood pressure interventions likely do not strongly affect memory, whether at the overall population level or in the population that would survive under both the natural course and the intervention (the always survivor stratum).
%\end{center}
%\maketitle
%\newpage

%%%%%%%%%%%%%%%%%%%%%%%%%%%%%%%%%%%%%%%%%%%%%%%%%%%%%%%%%%%%%%%%%%%%%%%%%%%%%%%%%%%%%%%%%%%%%%%%%%
\vspace{0.25cm}

%%%%%%%%%%%%%%%%%%%%%%%%%%%%%%%%%%%%%%%%%%%%%%%%%%%%%%%%%%%%%%%%%%%%%%%%%%%%%%
\newpage
\begin{center}
\textsc{1. Introduction}
\end{center}

Hypertension is a well-established risk factor for cognitive impairment and dementia \parencite{livingston2020}. However, the potential of promoting cognitive abilities through antihypertensive treatment remains poorly understood \parencite{iadecola2016}. Studying the link between hypertension and cognition is complicated by the temporal dimension, where cognitive alterations caused by hypertension are processes that likely unfolds over decades rather than years, making it challenging to conduct randomized clinical trials (RCTs). Although prospective studies can never replace RCTs, applying more sophisticated statistical methods/methodology to existing datasets could provide additional insights.

%%%%%%%%%%%%%%%%%%%%%%%%%%%%%%%%%%%%%%%%%%%%%%%%%%%%%%%%%%%%%%
%% Hypothetical interventions using the G-formula
The G-formula \parencite{robins1986new} can estimate effects of hypothetical interventions from longitudinal observational data with time-varying confounding (e.g. \citeauthor{zhou2020} \citeyear{zhou2020}; \citeauthor{taubman2009} \citeyear{taubman2009}). %Note, we use the term \textit{hypothetical intervention} to differentiate between an intervention from using prospective data (as in our case) and a clinical trial intervention. 
For a binary risk factor this often involve setting the risk factor to a fixed value (Yes/No) (e.g. \cite{zhou2020}), or by modifying a subject's odds of receiving treatment, i.e. a hypothetical intervention based on incremental propensity scores \parencite{kennedy2019}. 
However, for a continuous risk factor, setting the treatment dose to a fixed value for all subjects is often considered to result in highly unrealistic scenarios with limited practical use. For example, a hypothetical intervention where all subjects in the population fix their systolic blood pressure (sBP) levels at 140 mm Hg would not take into account the natural variation that is also seen among non-hypertensive subjects. Another complicating factor for a continuous risk factor arises when only subjects meeting a diagnostic criteria are offered treatment. For instance, only subjects with hypertension would be offered antihypertensive agents to control their blood pressure levels. 

Two approaches for hypothetical interventions on a continuous risk factor, which only intervene on subjects that do not reach a prespecified threshold, are the \textit{threshold intervention} \parencite{taubman2009} and the \textit{representative intervention} (e.g., \cite{picciotto2012, young2018}). The former assigns the threshold value to all individuals who do not meet the prespecified target, while the latter assigns a value drawn from the distribution among those in the population meeting the target level. Hence, both approaches ensure treatment remains within a prespecified interval. Nevertheless, it is common for a significant proportion of the population to be unable to control their sBP at optimal levels, due to factors such as poor response to medications. One alternative approach in these settings is an \textit{incremental intervention}, where a subject’s treatment is instead shifted downward (or upwards) by some prespecified function \parencite{haneuse2013, munoz2012}. However, their approaches are not valid for longitudinal data with dropout and deaths.

%%%%%%%%%%%%%%%%%%%%%%%%%%%%%%%%%%%%%%%%%%%%%%%%%%%%%%%%%%%%%%%
%% Mortal cohort inference
Attrition due to dropout and deaths is common in longitudinal aging studies, especially when subjects are followed for an extended duration. The G-formula can address both missing response and missing covariate data under an assumption of missing at random (MAR; \cite{robins1995analysis, young2018, kim2021}). Further strategies for sensitivity analyses have been proposed in situations where the missingness is thought to be missing not at random (MNAR) \parencite{josefsson2021bayesian, josefssonbiostatistics}.
However, studying cognition after death is typically not defined and is not of interest, leading to consideration of a \textit{mortal cohort} approach \parencite{dufouil2004}. Mortal cohort inferences truncate subjects’ outcomes after death. The first approach, \textit{partly conditional inference} (PCI), conditions on the sub-population that is still alive at the end of follow-up \parencite{kurland2005directly, wen2018methods, wen2018semi}, while the second approach, \textit{principal stratification}, conditions on the subpopulation that would survive through the end of follow-up regardless of treatment assignment, known as \textit{the always survivor stratum} \parencite{frangakis02, frangakis07}.
Although the two approaches have similarities, they differ with respect to the type of information they generate. PCI has been proposed for estimating non-causal associations and is particularly useful for prognostication, i.e., when the interest is in the health outcome among all individuals who are alive. This approach has, for example, been considered through the use of the G-formula in hypothetical interventions targeting risk factors for coronary heart disease \parencite{taubman2009}.
In contrast, principal stratification is useful in situations where survival differs between treated and control groups and has been applied when the interest is in etiological effect estimates (e.g. \cite{hayden2005, tchetgen2014identification, wang2017deaths}). 

For observational data with time-varying confounding and possibly MNAR missingness, \citeauthor{shardell2018joint} (\citeyear{shardell2018joint}) proposed a parametric shared parameter model with g-computation to identify a survival average causal effect (SACE). \citeauthor{josefsson2021bayesian} (\citeyear{josefsson2021bayesian}) proposed a g-computation approach for the SACE, using Bayesian semi-parametric modeling. \citeauthor{tan2021SACE} (\citeyear{tan2021SACE}) took a different approach to identifying the SACE, where they impute counterfactual survival status of each subject and then compute a SACE for the always survivor stratum using marginal structural models. However, a drawback with many approaches is that monotonicity is required for valid inference, i.e. for a beneficial treatment, individuals surviving under treatment would also survive in the absence of treatment, which may be too strong of an assumption in many settings. To overcome this issue, \citeauthor{zaidi2019} (\citeyear{zaidi2019}) provide methods for sensitivity analyses when the standard monotonicity assumption does not hold, and additionally provide a population level characterization of such causal eﬀects. In a clinical trials setting, \citeauthor{roy2008principal} (\citeyear{roy2008principal}) and \citeauthor{lee2010causal} (\citeyear{lee2010causal}) also provided methods for sensitivity analyses, utilizing a less stringent \textit{stochastic monotonicity} assumption. However, their approaches are not valid for prospective data with a time-varying treatment and time-varying confounding.

%%%%%%%%%%%%%%%%%%%%%%%%%%%%%%%%%%%%%%%%%%%%%%%%%%%%%%%%%%%%%%
% Our study
%%%%%%%%%%%%%%%%%%%%%%%%%%%%%%%%%%%%%%%%%%%%%%%%%%%%%%%%%%%%%%
In the present study we investigate population effects on episodic memory of an sBP intervention using prospective cohort data from a large representative sample \parencite{nyberg2020betula}. An important limitation in prior research is the lack of sufficient consideration of altered mortality rates when evaluating the potential risk reduction. As such, we propose both partly conditional inference and inference on the always survivor stratum. Thus, our study extends previous research by providing both etiological and prognostic intervention effect estimates, which are both critical for public health practice, offering a more comprehensive understanding of the intervention's impact. For estimation, we propose a Bayesian semi-parametric approach based on the extended G-formula \parencite{robins2004}. In addition, we introduce a novel sparsity-inducing Dirichlet hyperprior designed for regression with longitudinal data, and demonstrate the usefulness of our approach in a simulation relative to other Bayesian tree ensemble approaches. 

The paper is organized as follows. In Section 2, we introduce the Betula data. In Section 3, we introduce the proposed intervention in a Bayesian setting. In Section 4, we describe the two types of mortal cohort inferences. In Section 5, we provide computational details and introduce the novel sparsity-inducing Dirichlet hyperprior for longitudinal data. In Section 6, we conduct a simulation study to assess the operating characteristics of the new prior, and in Section 7, we apply the methodology to the Betula data to answer the substantive question of interest. We close in Section 8 with some conclusions.

%%%%%%%%%%%%%%%%%%%%%%%%%%%%%%%%%%%%%%%%%%%%%%%%%%%%%%%%%%%%%%%%%%%%
%%%%%%%%%%%%%%%%%%%%%%%%%%%%%%%%%%%%%%%%%%%%%%%%%%%%
\begin{center}
\textsc{2 A blood pressure intervention}
\end{center}
%\section{A blood pressure intervention}
The subjects in this study are participants in the Betula study, a population-based prospective cohort study initiated in 1988-1990, with the objective to study memory, aging and dementia \parencite{nyberg2020betula}. We consider 15-years of longitudinal data collected over four time points, with data collection occurring at approximately five-year intervals. A total of n=800 participants were included in the current application, with 100 participants from each of the eight age cohorts $35, 40, \ldots , 70$ years old at enrollment. The cognitive outcome in this study was assessed using a composite episodic memory task, with a range of 0 to 76 (\cite{josefsson12}). We included a set of baseline confounding factors, including age at enrollment, sex, and years of education, as well as a set of time-varying cardiovascular risk factors: body mass index (BMI), smoking, serum-cholesterol, and self-reported type 2 diabetes, as well as previous episodic memory measures. See Table \ref{betula_demographics} for baseline summary statistics. 38\% of participants dropped out of the study before the 15-year follow-up, and among these, 18\% died or developed dementia.

Although the recommended sBP levels have varied over time, we define hypertension as a sBP measure greater than 140 mm Hg. In Section 3 we propose an \textit{incremental threshold intervention}. The intervention loosely imply that, at each follow-up time point, sBP levels is shifted downwards by a prespecified value in individuals with hypertension, while no intervention is imposed on individuals without hypertension.

%%%%%%%%%%%%%%%%%%%%%%%%%%%%%%%%%%%%%%%%%%%%%%%%%%%%
\begin{center}
\textsc{3 Methods}
\end{center}
%\section{Methods}
%This section introduces the data structure, notation and the proposed hypothetical incremental intervention in a Bayesian setting. 
%\begin{center}
%\textit{3.1 Data structure and notation}
%\end{center}
%\subsection{Data structure and notation}
Consider a prospective cohort study where measurements on a set of variables are collected on each of $i = 1, 2, \ldots, n$ subjects over a specified follow-up period at $t=1, 2, \ldots,T$ fixed time points. Let $L_{it}$ denote the vector of time-varying confounders, and $A_{it}$ denote a continuous treatment, measured at time $t$. Denote the continuous outcome measure by $Y_{it}$, where interest in this study is the outcome measure at the last time point $T$. Furthermore, let $S_{it}=1$ indicate if a subject is alive at time point $t$; and $S_{it}=0$ otherwise. Let $R_{it}=1$ indicate if subject $i$ has completed the cognitive testing at time $t$; and $R_{it}=0$ otherwise. We have monotone missingness, so if $R_{it}=0$, then $R_{ik}=0$ for $k > t$. We let $T^r_i$ denote the number of time points a subject participates in the study, and $T^s_i\geq T^r_i$ denotes the number of time points the subject is alive. Let the observed history of a time-varying variable be denoted with an overbar and the future with an underbar. %For example, $\bar{L}_{it}=(L_{i1},L_{i2},\ldots,L_{it})$ denote the confounder history for individual $i$ through test wave $t$ and $\ubar{A}_{it}=(A_{it},A_{it+1}, \ldots,A_{iT})$ denote the future treatment values. 
We also write $H_{i}$ to denote all the observed past history just prior to the outcome variable at the last time point $T$, such that $H_{i}=(\bar{Y}_{it-1}=\bar{y}_{it-1}, \bar{A}_{it}=\bar{a}_{it}, \bar{L}_{it}=\bar{l}_{it}, \bar{R}_{it}=\bar{r}_{it}, \bar{S}_{it}=1)$, with support on the set of possible histories $\mathcal{H}_{i}$ and where lower-case letters denote possible realizations of a random variable.
At each time point $t$, we assume data is observed in the temporal ordering $S_{it}, R_{it}, L_{it}, A_{it}, Y_{it}$, and for each subject we observe $(\bar{s}_{iT^s_i}, \bar{r}_{iT^r_i}, \bar{l}_{iT^r_i}, \bar{a}_{iT^r_i}, \bar{y}_{iT^r_i})$. In what follows, we suppress the subscript $i$ to simplify notation.

\begin{center}
\textit{3.1 A hypothetical incremental threshold intervention}
\end{center}

Consider a scenario where only subjects meeting a classification or diagnostic criteria would be recommended for treatment. In this setting, a hypothetical intervention that depends on the natural value of treatment (NVT) is a viable alternative. Following \citeauthor{young2014} (\citeyear{young2014}), the NVT is defined at wave $t$ as the treatment measure that would have been observed at time $t$ if the intervention discontinued right before $t$. We denote the NVT and the intervention regime by $A_{t}^*$ and $g^{int^*}$ respectively, and let the conditional intervention density under an intervention based on the NVT be characterized by $p^{int^*}(a_{t} | y_{t-1}, a_{t}^*, \bar{a}_{t-1}, \bar{l}_{t}, \bar{r}_{t}, s_{t}=1)$. 
In the present study, we only intervene on subjects whose NVT at time $t$ does not meet a particular threshold, here denoted by $\tau$. Specifically, for individuals whose NVT is greater than $\tau$ at time $t$, we assign a new treatment value by shifting the NVT downwards by some value $\delta_t$. We further let $\delta_t$, i.e. the shift, correspond to a parameter, and quantify the uncertainty in the intervention by imposing a prior on $\delta_t$. The prior reflects the researchers beliefs about the possible values for treatment to take under the hypothetical intervention. As such, treatment is assigned using the prior, $p^{\delta_t}(\cdot)$. Specifically, if $a_t^* \leq \tau$, then $p^{int^*}(a_{t} | \bar{y}_{t-1}, a_{t}^*, \bar{a}_{t-1}, \bar{l}_{t},\bar{r}_{t}, \bar{s}_{t}=1) = 1,$
if $a_t=a_t^*$.  However, if $a_t^* > \tau$, then 
$p^{int^*}(a_{t} | \bar{y}_{t-1}, a_{t}^*, \bar{a}_{t-1}, \bar{l}_{t},\bar{r}_{t}, \bar{s}_{t}=1) = p^{\delta_t}(\delta_t) I\{a_t=a_t^* - \delta_t\}.$ 

%%%%%%%%%%%%%%%%%%%%%%%%%%%%%%%%%%%%%%%%%%%%%%.%%%%%%%%%%%%%%%%%%%%%%%%%%%%%%%%%
\begin{center}
\textsc{4 Intervention effects for mortal cohorts}
\end{center}
%\section{Intervention effects for mortal cohorts}
In this section we describe two types of mortal cohort inferences under the proposed incremental intervention on the NVT: \textit{partly conditional inference} and \textit{inference on the the always survivor stratum}. For simplicity, we temporarily restrict our description to the regimes $g\in\{g^{int^*}, g^{nc}\}$, where $g^{int^*}$ is a deterministic intervention regime based on the NVT, with support on the set of possible regimes $\mathcal{G}$, and $g^{nc}$ denotes the natural course, i.e. the natural progression in the presence of no intervention. Let $\bar{Y}_T(g)$, $\bar{A}_T(g)$, $\bar{R}_T(g)$, $\bar{S}_T(g)$, and $\bar{L}_T(g)$ represent the counterfactual outcome, treatment, missingness, survival, and covariate histories associated with the regime $g$. 
\begin{center}
\textit{4.1 Partly conditional inference}
\end{center}
%\subsection{Partly conditional inference}
Partly conditional models have been proposed to address truncation by death under hypothetical interventions. For these models inference is conditional on the sub-population alive at a specific time-point, in our application being alive at end of follow-up, i.e. time $T$. Hence, inferences focus on what we call a \textit{partly conditional intervention effect}: $PCIE = E[Y_T(g^{int^*}) | S_T(g^{int^*})=1] - E[Y_T(g^{nc}) | S_T(g^{nc})=1]$.
Although the $PCIE$ gives a crude comparison of differences in potential outcomes under the two regimes, which is useful for prognostication, it often implicitly compares potential outcomes for different underlying populations. For example, it is likely that an sBP intervention may affect survival rates; as such, the population surviving under the natural course will differ from the one surviving under the intervention.

\begin{center}
\textit{4.2 Assumptions and identification of the PCIE}
\end{center}
%\subsubsection*{Identifying assumptions and the extended g-formula} 
We now define a set of assumptions to identify the $PCIE$. 
%\begin{enumerate}

\textbf{C1} \textit{Consistency:} For $g\in(g^{int^*}, g^{nc})$ and $t=1,\ldots,T$, if $\bar{A}_t = \bar{A}_t(g)$, then $\bar{Y}_t = \bar{Y}_t(g)$, $\bar{R}_t=\bar{R}_t(g)$, $\bar{S}_t=\bar{S}_t(g)$ and $\bar{L}_t = \bar{L}_t(g)$.

\textbf{C2a} \textit{Conditional exchangeability under the natural course:} For $g^{nc}$ and $t=1,\ldots,T$, conditional on the history, the treatment and the potential outcome are independent. That is $\ubar{Y}_t(g^{nc})  \perp\!\!\!\perp A_t | \bar{Y}_{t-1}=\bar{y}_{t-1}, \bar{A}_{t-1}=\bar{a}_{t-1}, \bar{R}_{t}=\bar{r}_{t}, \bar{S}_t=1, \bar{L}_t=\bar{l}_t$ and \\ 
$\ubar{S}_{t+1}(g^{nc}) \perp\!\!\!\perp A_t | \bar{Y}_{t}=\bar{y}_{t}, \bar{A}_{t-1}=\bar{a}_{t-1}, \bar{R}_{t}=\bar{r}_{t}, \bar{S}_t=1, \bar{L}_t=\bar{l}_t$.

\textbf{C2b} \textit{Conditional exchangeability for NVT interventions:} For $g^{int^*}$ \\ $\left( \ubar{Y}_t(g^{int^*}), \ubar{A}_t^*(g^{int^*}) \right)  \perp\!\!\!\perp A_t | \bar{Y}_{t-1}=\bar{y}_{t-1}, \bar{A}_{t-1}=\bar{a}_{t-1}, \bar{R}_{t}=\bar{r}_{t}, \bar{S}_t=1, \bar{L}_t=\bar{l}_t$ and $\left( \ubar{L}_{t+1}(g^{int^*}), \ubar{S}_{t+1}(g^{int^*}) \right) \perp\!\!\!\perp A_t | \bar{Y}_{t}=\bar{y}_{t}, \bar{A}_{t-1}=\bar{a}_{t-1}, \bar{R}_{t}=\bar{r}_{t}, \bar{S}_t=1, \bar{L}_t=\bar{l}_t$. \\ 
This assumption states that under the intervention, the NVT at time $t$ has no direct effect on the cognitive outcome, or survival, except through future measurements of the treatment. An implication of this assumption is that the NVT is not itself a time-varying confounder \parencite{richardson2013, young2014}.

\textbf{C3} \textit{Sequential positivity:} If  $p(\bar{y}_{t-1}, \bar{a}_{t-1}, \bar{l}_{t}, \bar{r}_{t}, \bar{s}_{t} = 1) \neq 0$ then $p(a_{t} | \bar{y}_{t-1}, \bar{a}_{t-1}, \bar{l}_{t}, \bar{r}_{t}, \bar{s}_{t} = 1) > 0$ for all $\bar{a}_{t}$ consistent with regime $g^{int^*}$. 

\textbf{C4} \textit{Missingness at follow-up:} For the outcome, treatment and time-varying confounders, we assume missing at random conditional on survival (MARS). In particular, for all $t>1$, $p(y_t | \bar{a}_{t}, \bar{l}_{t}, r_t=0, \bar{s}_{t}=1, \bar{y}_{t-1}) = p(y_t | \bar{a}_{t}, \bar{l}_{t}, \bar{r}_{t}=1, \bar{s}_t=1, \bar{y}_{t-1})$, and similarly for $a_t$ and $l_t$. %$p(a_t | \bar{y}_{t-1}, \bar{a}_{t-1}, r_t=0, \bar{s}_{t}=1, l_{t}) = p(a_t | \bar{y}_{t-1}, \bar{a}_{t-1}, r_t=1, \bar{s}_{t}=1, l_{t})$, and $p(l_t | \bar{y}_{t-1}, \bar{a}_{t-1}, r_t=0, \bar{s}_{t}=1, l_{t-1}) = p(l_t | \bar{y}_{t-1}, \bar{a}_{t-1}, r_t=1, \bar{s}_{t}=1, l_{t-1})$. 
%\end{enumerate}

%\textit{Identification of the $PCIE$.}
%\subsubsection*{Identification of the $PCIE$}
For longitudinal cohort data, under the assumptions C1, C2a, C3 and C4, the mean outcome at time $T$ under the natural course, $g^{nc}$, can be computed using the g-formula (Robins, 1986) conditioning on survival, %That is %$E_{H_{nc}}\{E[Y_T | S_T=1, H_{nc}]\} = $
\begin{align}\label{gformula}
& \int_{\bar{s}_{T}}\int_{\bar{r}_{T}}\int_{\bar{l}_{T}}\int_{\bar{a}_{T}}\int_{\bar{y}_{T-1}} E[Y_T  | \bar{y}_{T-1}, \bar{a}_{T}, \bar{l}_T, \bar{r}_{T}, \bar{s}_{T}=1] \times \prod_{k=1}^T p(a_k | \bar{y}_{k-1}, \bar{a}_{k-1}, \bar{l}_{k}, \bar{r}_{k}, \bar{s}_{k}=1) \nonumber \\ 
& \quad \times p(l_k | \bar{y}_{k-1}, \bar{a}_{k-1}, \bar{l}_{k-1}, \bar{r}_{k}, \bar{s}_{k}=1) \times p(r_k | \bar{y}_{k-1}, \bar{a}_{k-1}, \bar{l}_{k-1}, \bar{r}_{k-1}, \bar{s}_{k}=1) \nonumber \\ 
& \quad \times p(s_k | \bar{y}_{k-1}, \bar{a}_{k-1}, \bar{l}_{k-1}, \bar{r}_{k-1}, \bar{s}_{k-1}=1) \times p(y_{k-1} | \bar{y}_{k-2}, \bar{a}_{k-1}, \bar{l}_{k-1}, \bar{r}_{k-1}, \bar{s}_{k-1}=1). %\nonumber \\
%& \qquad d\bar{y}_{T-1} d\bar{a}_T d\bar{l}_T d\bar{r}_T d\bar{s}_T.
\end{align}
For the Betula data we perform sensitivity analyses to investigate robustness of results to violations of assumptions C2, C3, and C4, (see Section 7).

In contrast, for the intervention, $g^{int^*}$, under assumptions C1, C2b, C3 and C4, the mean outcome at time $T$ can be computed using the \textit{Extended g-formula} \parencite{robins2004}. The key difference is that $p(a_k | \bar{y}_{k-1}, \bar{a}_{k-1}, \bar{l}_{k}, \bar{r}_{k}, \bar{s}_{k}=1)$ in (\ref{gformula}) is replaced by $\int_{a_{t}^*} p^{int^*}(a_k | \bar{y}_{k-1}, a^*_{k}, \bar{a}_{k-1}, \bar{l}_{k}, \bar{s}_{k}=1) \times p(a^*_k | \bar{y}_{k-1}, \bar{a}_{k-1}, \bar{l}_{k}, \bar{s}_{k}=1)$; details are given in Supplementary Appendix A. 
Similar to (\ref{gformula}), Monte Carlo integration is implemented over the history under the intervention, denoted by $H_{int^*}$, among those who are alive. However, instead of sampling pseudo treatment data from the observed data distribution, as in (\ref{gformula}), pseudo NVT data is first sampled from the observed data distribution. For those whose NVT does meet the threshold $\tau$, treatment is set to the NVT, and otherwise, treatment is assigned such that the NVT is shifted downwards by $\delta_t$, as described in Section 3.2. 

The $PCIE$ is then computed as the differences in expected outcomes under the two regimes $g^{nc}$ and $g^{int^*}$. More details are given in Supplementary Appendix B.

%%%%%%%%%%%%%%%%%%%%%%%%%%%%%%%%%%%%%%%%%%%%%%%%%%%%%%%%%%%%%%%%%%%%%%%%%%%%%%%%%%%%%%%%%%
\begin{center}
\textit{4.3 Inference on the always survivor stratum}
\end{center}
%\subsection{Inference on the always survivor stratum}

Conditioning on survival as described in Section 4.1 may introduce bias due to the fact that survival is a post-randomization event \parencite{zhang03}. Hence, the partly conditional approach in Section 4.2 cannot address etiological research questions in mortal cohorts. An alternative estimand is the causal effect on the subpopulation who would survive through the end of follow-up regardless of treatment regime (\cite{frangakis02, frangakis07}). Our goal here is to estimate the difference in expected potential outcomes between $g^{nc}$ and $g^{int^*}$ in the always survivor stratum. That is, what we call a \textit{survivors average intervention effect}: $SAIE = E[Y(g^{int^*}) - Y(g^{nc}) \mid S_T(g^{nc})=S_T(g^{int^*})=1]$. Note, previous studies have shown a the link between reduced sBP and reduced all-cause mortality \parencite{bundy2017}. Hence, it is likely that an intervention targeting sBP not only may have an effect on cognition but also on mortality rates. In such situations, the $PCIE$ would compare potential outcomes for different underlying populations. In particular, if subjects live longer under the risk-reducing intervention this population may include less healthy subjects who would die under the natural course, masking a possible intervention effect.
The $SAIE$ on the other hand, compare potential outcomes from the same population, i.e. a population consisting of subjects who would survive both when intervening on sBP and under the natural course. As such, it provides an estimate that adjusts for differences in mortality rates between the two regimes.
In our application we are interested in both a prognostic effect estimate (i.e. the $PCIE$) and an etiological effect estimate (i.e. the $SAIE$). 

For identification of the $SAIE$, we need two additional assumptions. 

\textbf{C5} \emph{Stochastic Monotonicity:} $\Pr[S_{T}(g^{int^*}) = 1 \mid S_{T}(g^{nc}) = 1] \geq \Pr[S_{T}(g^{int^*}) = 1 \mid S_{T}(g^{nc}) = 0]$. In the current application, we impose an intervention that improves individuals’ health. As such, we assume the probability of survival under the intervention $g^{int^*}$ is higher among those subjects that would survive under the natural course $g^{nc}$, compared to those who would not. We denote $\Pr[S_T(g) =1]$ by $\psi^g$ for $g$ in $\{g^{nc},g^{int^*}\}$, and we assume $\psi^{g^{int^*}} \geq \psi^{g^{nc}}$. We further assume $\Pr[S_T(g^{int^*}) = 1 \mid  S_T(g^{nc}) = 1] =  \psi^{g^{int^*}} + \lambda \left[\min\left\{1, \frac{\psi^{g^{int^*}}}{\psi^{g^{nc}}}\right\} - \psi^{g^{int^*}}\right]$, where $\lambda$ is a sensitivity parameter \parencite{roy2008principal, lee2010causal}. Note, $0\leq \lambda \leq 1$. If $\lambda=1$ this assumption corresponds to a \textit{deterministic monotonicity} assumption where subjects who were to be alive under the observed regime $g^{nc}$ would also be alive under the intervention $g^{int^*}$, i.e. $S_{T}(g^{nc}) \leq S_{T}(g^{int^*})$. In contrast, if $\lambda=0$ this corresponds to an assumption of $S_{T}(g^{nc})$ and $S_{T}(g^{int^*})$ being independent, which we in the current application expect to be unlikely.

\textbf{C6} \emph{Difference in expectations of outcomes when comparing different strata}. For $g\in\{g^{int^*}, g^{nc}\}$, $g'\in\{g^{int^*}, g^{nc}\}$ and $g \neq g'$, we assume, $E[Y_T(g) | S_T(g') = S_T(g)= 1] - E[Y_T(g) | S_T(g) = 1, S_T(g') \neq 1] = \Delta^g$, where $\Delta^g$ is a sensitivity parameter. We further assume that $\Delta^g=\Delta^{g'}=\Delta$, i.e. that the difference in expectations of potential outcomes when comparing different strata is the same for the two contrasting regimes. Note, if $\Delta=0$ there is no difference in potential outcomes when comparing the always survivor strata to the strata where individuals were to live under regime $g$ but not under the contrasting regime $g'$. In contrast, if $\Delta>0$ this implies a higher potential outcome is observed for the always survivor stratum. In the current application we expect higher memory scores, i.e. a better memory, for the always survivor stratum.

The identification details of the $SAIE$ are given in Supplementary Appendix B. Briefly, under assumptions C1-C6, the $SAIE$ is given by: \\
$PCIE + \Delta \left\{ \psi^{g^{int^*}} + \lambda \left(U - \psi^{g^{int^*}}\right) \right\} \left(1 -  \frac{1}{U} \right)$, 
where $U=\min\left\{1, \frac{\psi^{g^{int^*}}}{\psi^{g^{nc}}}\right\}$. The equation represents a location shift upwards from the $PCIE$ for a beneficial intervention where we expect the participants to improve their health under the intervention compared to the natural course.  
Note, $\psi^{g} = E_{H^{g}}\left\{\Pr[S_T=1| H^{g}]\right\}$, for $g\in\{g^{int^*}, g^{nc}\}$, are computed by marginalizing over the distributions of the set of temporally preceding variables $H^{nc}$ using the g-formula and $H^{int^*}$ using the extended g-formula. 

%%%%%%%%%%%%%%%%%%%%%%%%%%%%%%%%%%%%%%%%%%%%%%%%%%%%%%%%%%%%%
\begin{center}
\textsc{5 A semi-parametric approach for estimation}
\end{center}
%\section{A semi-parametric approach for estimation}
In this section, we develop a Bayesian semi-parametric (BSP) modeling approach using \textit{soft} Bayesian additive regression trees (BART) models to estimate each of the conditional distributions in (\ref{gformula}) and (\ref{extgformula}) in Supplementary appendix A. The \textit{soft} BART model improves upon the traditional \textit{hard} BART model by addressing the lack of smoothness inherent in ensemble approaches. To enhance valid inference, especially in scenarios with multiple time points and a large set of regressors, we extend the current \textit{soft} BART algorithm by introducing a novel sparsity-inducing hyperprior for longitudinal data. This longitudinal Dirichlet prior promotes parsimony by penalizing groups of predictors less likely to be important in longitudinal prediction settings. We refer to our model as a longitudinal Dirichlet additive regression trees (LDART) approach.

As for traditional (\textit{hard}) BART models, we have that the mean function of a variable conditional on the history, is given by the sum of regression trees, $\mu_{t} = \sum_{b=1}^B g_{y_{t}}(\mathcal{T}_b, \mathcal{M}_b)$, where $\mathcal{M}_b$ is the parameters for the leaf nodes and $\mathcal{T}_b$ the tree structure. In general, the distribution of the continuous variable are moreover specified as normal, and the binary variables are specified as probit models.

The regularization prior, controlling the complexity of the sum-of-trees, differ in two ways between the standard BART algorithm and \textit{soft} BART. First, for \textit{soft} BART, the path down the decision tree is probabilistic, where the sharpness of the decision is controlled by the bandwidth parameter. Thereby adapting to the smoothness level of the regression function even for low dimensional tree ensembles, a known shortcoming for \textit{hard}) BART models. 
Second, the $j$th predictor, from the set of $P$ predictors, is chosen according to a probability vector $\mathbf{q} = (q_1, \ldots, q_P)$, typically specified as uniform. In contrast, \textit{soft} BART models incorporate a Dirichlet distribution for $\mathbf{q}$, such that $\mathbf{q} \sim \mathcal{D}\left( \alpha/P, \ldots, \alpha/P \right)$. The sparsity-inducing prior was first introduced for high dimensional settings \parencite{linero2018DART}, and can be extended to allow sparsity both within and between groups \parencite{linero2018SoftBart, du2019grouping}. 
Determining the degree of sparsity in the model is done by placing a prior on $\alpha$, as such, $\frac{\alpha}{\alpha + \rho} \sim Beta\left(a, b \right)$. Default specifications for $a, b$ and $\rho$ is $0.5$, $1$ and $P$ respectively. 
This sparsity inducing prior is implemented for both \textit{hard} and \textit{soft} BART. We refer to BART models with this prior as Dirichlet additive regression trees (DART). 

\begin{center}
\textit{5.2 A grouping prior for longitudinal data}
\end{center}
%\subsection{A grouping prior for longitudinal data}
In a longitudinal setting, one may want to introduce a larger degree of sparsity for predictors collected at earlier time points, if they are thought to be less strongly associated with the response. Below we introduce a novel sparsity-inducing prior. 

We consider a hierarchical structure where the predictors are divided into two groups: predictors measured at the \textit{current} time point $t$ and predictors measured at \textit{past} time points $k=1,\ldots,t-1$. %Note, $t$ refer to the time point when the \textit{current} response (in the g-formula) is measured. 
Moreover, baseline predictors, e.g. age and sex, are grouped with the \textit{current} predictors into one group, since we want to impose the same degree of sparsity on these predictors. 
Let $w$ refer to the probability of a predictor being from the set of \textit{current} predictors, versus from the set of \textit{past} predictors; we specify a beta prior, $w \sim Beta\left(a, b \right)$. For the \textit{current} predictors, let $v^t_{j}$ refer to the probability of selecting the $j$th predictor from the set of $P_t$ \textit{current} predictors $\textbf{v}^t = (v^t_{1},\ldots,v^t_{P_t}$); $\textbf{v}^t$ is given a Dirichlet prior $\textbf{v}^t \sim \mathcal{D}\left( \eta/P_t, \ldots, \eta/P_t \right)$. So a priori, the probability of selecting the $j$th \textit{current} predictor to construct a given split is equal to $w \times v^t_{j}$.
The \textit{past} predictors are divided into $t-1$ groups, where $P_{k}$ is the size of group associated with time $k$. Let $u^k$ denote the probability of selecting the $k$th set of time-varying predictors, and let $v^k_{j}$ denote the probability of selecting the $j$th predictor in the $k$th set of time-varying predictors. To introduce sparsity related to proximity in time to the response we give $\textbf{u}$ a grouping Dirichlet prior $\textbf{u} \sim \mathcal{D}\left( \alpha_1/t-1, \ldots, \alpha_{t-1}/t-1 \right)$, where we determine the degree of sparsity in the model by placing a prior on $\alpha_k$ such that a larger degree of sparsity is obtained for smaller $k$'s (see details below). Similarly $\textbf{v}^k$ is also given Dirichlet prior $\textbf{v}^k \sim \mathcal{D}\left( \varphi^k/P_k, \ldots, \varphi^k/P_k \right)$. So a priori, the probability of selecting the $j$th predictor from the $k$th time point to construct a given split is equal to $(1-w) \times u^k \times v^k_{j}$.

%\textit{Hyperpriors.}
%\subsubsection{Hyperpriors}
For the hyperparameters $\eta$ and $\varphi^k$ we use the default Beta hyperprior as implemented in DART. 
For $\alpha_j$ we specify a Beta hyperprior $\frac{\alpha_j}{\alpha_j + \rho_j} \sim Beta\left(c_j, d_j \right)$; we let $c_j$ vary as a linear function of time, set $d_j=1$, and $\rho_j=t$ for all $j$. To introduce sparsity related to proximity in time to the response we set $c_j = 1-\frac{(t - j)}{t-1} \times 0.5$. This results in ordered hyperpriors for the $\alpha_j$s, where $c_1=0.5$ for predictors measured at the first time point and $0.5<c_{j-1}<c_j<1$ for $j>1$. That is, a higher degree of sparsity is introduced for predictors measured earlier in time and less sparsity for predictors measured closer in time to the response.
The simulations in Section 6 shows the strong performance of these hyperprior choices. Details about the posterior computations of the L-Dirichlet (LDART) prior are obtained in the Supplementary Appendix C.

\begin{center}
    \textit{5.3 Algorithm to compute the $PCIE$ and $SAIE$}
\end{center}
%\subsection{Algorithm to compute the $PCIE$ and $SAIE$}
The $PCIE$ and the $SAIE$ can be computed using the algorithm described in the Supplementary Appendix C and is implemented in the R package \textit{GcompLDART} (available upon request). 

%%%%%%%%%%%%%%%%%%%%%%%%%%%%%%%%%%%%%%%%%%%%%%%%%%%%%%%%%%%%%
\begin{center}
    \textsc{6 Simulation study}
\end{center}
%\section{Simulation study}
In this simulation study we evaluate the performance of LDART and compare it to other BART specifications. 
%\begin{center}
%    \textit{6.1 Data generating process and simulation scenarios}
%\end{center}
%\subsection{Data generating process and simulation scenarios}
Data were generated according to a modified version of the Friedman's example \parencite{friedman91}, mirroring a longitudinal study with five time-varying predictors and a continuous outcome, with $f(\bar{\mathbf{x}}_{i}) = \sum_{t=1}^T c_t  [10 \mathrm{sin}(\pi  x_{t}^1 x_{t}^2) + 20  (x_{t}^3 - 0.5)^2 + 10  x_{t}^4 + 5 x_{t}^5]$, where $Y_{iT} \sim \mathrm{Normal}\left\{ f(\bar{\mathbf{X}}_{i}), \sigma_T\right\}$, $T=4$, $\sigma_T=10$, and $c_t$ determines the strength of association between the predictors at time $t$ and the outcome. 

The longitudinal predictors, $\bar{\mathbf{X}}_{i} = \bar{\mathbf{X}}_i^1, \ldots, \bar{\mathbf{X}}_i^T$, were generated as correlated uniform random variables, generated as multivariate normal variables, with zero mean and covariance as described below, and then transformed to uniform variables. As such, $\bar{X}_{it}^p \sim Uniform(0,1)$, where the between time-points covariance for predictor $X^p$ is $Var(X^p_{it},X^p_{ik})=d_{tk}^2$. Here, we specified $d_{jj-1}=0.4$, $d_{jj-2}=0.2$ and $d_{jj-3}=0.1$, reflecting a stronger correlation for measures closer in time. All correlations between different predictors were set to zero. 
We consider missingness in the form of dropout and assumed approximately $15\%$ dropped out at each follow-up time point $t=2,\ldots,T$. As such, dropout at time $j$ was generated as, $\pi_j(\bar{\mathbf{x}}_{i}) =  h_j \sum_{t=1}^{j-1} [x_{it}^1 + x_{it}^2 + x_{it}^3 + x_{it}^4 + x_{it}^5]$, where $R_{ij} \sim \mathrm{Binomial}\left\{ \pi_j(\bar{\mathbf{x}}_{i}) \right\}$ and $\bar{h}=(0.75, 0.4, 0.25)$. 

We simulated 650 datasets of size $n = \left(250, 750, 1500\right)$, and four different scenarios by varying $\bar{c}=\left(c_T,\ldots,c_1\right)$: \textit{Scenario A}: $\bar{c}=(1,0.75,0.5,0.25)$, reflecting a decreasing, although a relatively strong, association over time between the predictors and the outcome; \textit{Scenario B}: $\bar{c}=(1,0.75,0.5,0.25)^3$, i.e. a sharp decrease in association over time; \textit{Scenario C}: $\bar{c}=(1,0,0,0)$, i.e. only predictors at the closest time point are associated with the outcome; \textit{Scenario D}: $\bar{c}=(0.25,0.25,0.25,0.25)$, i.e. a weak and stable association with the outcome for all time points. For all scenarios we estimated $E_{H^{g^{nc}}}\{E[Y_T(g^{nc}) \mid H^{g^{nc}}]\}$ using equation (\ref{gformula}). 

We compared the performance of our proposed approach (LDART) with two \textit{hard} BART models: the standard BART algorithm (HBART) and BART with the sparsity-inducing Dirichlet prior (HDART) (as implemented in the R package BART), as well as two \textit{soft} BART models: the standard SoftBart algorithm (SDART; as implemented in the R package SoftBart) and SoftBart incorporating only the predictors from the last time point in the model (SDART-lag1). We specified 50 regression trees for each model. To evaluate their performance, we compared relative absolute bias, root-mean-squared error (RMSE), and 95\% coverage probabilities (CPs).

The models use their respective default priors (if not stated otherwise) and used 1200 warm-up iterations, 2000 sampling iterations, a thinning of 4, and 48 chains. All simulations were done using R-4.1 at the \textit{High Performance Computing Center North} (HPC2N) and code is available from the author upon request.

\begin{center}
    \textit{6.1 Simulation Results}
\end{center}
%\subsection{Simulation Results}
The simulation results are presented in detail in Table \ref{table_sim_study_1}. For all scenarios and models, the absolute relative biases and RMSEs are smaller for the larger sample sizes, as expected. Moreover, the coverage probabilities are close to the nominal values for most scenarios, i.e. CPs: $0.91 - 0.98$.
The \textit{soft} BART models (including LDART) using the full set of predictors, generally have the smallest biases and RMSEs. LDART improves on SDART in all scenarios except one (Scenario A and for n=1500), where SDART is slightly better than LDART. 
In Scenario B and C, (that have a strong association between predictors at the last time point and the outcome and weak or zero with previous time points), the \textit{soft} BART model using only the last time-point predictors (SDART-lag1) performs similarly as the other \textit{soft} BART models in terms of bias, RMSE, and CPs. For the other two scenarios, SDART-lag1 has higher biases and RMSEs, and smaller CPs ($0.91 - 0.95$). 
The two \textit{hard} BART models produces biases and RMSEs that are notably larger than the other models. In addition, the standard \textit{hard} BART model produces overly conservative credible intervals in many scenarios, i.e. CPs $\geq 0.97$. 

%%%%%%%%%%%%%%%%%%%%%%%%%%%%%%%%%%%%%%%%%%%%%%%%%%%%%%%%%%%%%

%%%%%%%%%%%%%%%%%%%%%%%%%%%%%%%%%%%%%%%%%%%%%%%%%%%%%%%%%%%%%
\begin{center}
    \textsc{7 Analysis of the Betula data}
\end{center}
%\section{Analysis of the Betula data}
We consider an intervention where we assign sBP using the \textit{incremental threshold intervention} described in Sections 2 and 3. If a subjects’s NVT at times $t=2,\ldots,T$ is greater than $\tau=140$, then sBP is assigned by imposing a incremental shift on the NVT, and otherwise we do not intervene.
We specify a triangular prior for the shift parameter $\delta_t$, such that $\delta_t \sim Triangular(L_{\tau}, 0, L_{\tau})$. The lower bound and mode are given by $L_{\tau}\approx 44$ mm Hg, corresponding to two standard deviations of baseline sBP, and the upper bound is 0. The intervention implies a negative shift from the NVT with between 0 and 44 mm Hg for individuals with a $NVT>140$ at each follow-up time point.

To estimate the $SAIE$, we also specify priors on the sensitivity parameters in Assumptions C5 and C6. We specify triangular priors for the sensitivity parameters $\Delta$ and $\lambda$. For $\Delta$, we assume $\Delta \sim \mathrm{Triangular}(0, U_{\Delta}, U_{\Delta}),$ with upper bound (and mode) set to $4.6\%$; $4.6\%$ corresponds to the estimated difference in baseline memory between those who are alive at time point $T$ and those who are dead, adjusting for age and sex. It is unlikely the difference between the strata would be bigger than this. We further assume $\lambda \sim Triangular(0.5, 1, 1)$, where the upper bound corresponds to the deterministic monotonicity assumption while the lower bound reflects a moderate correlation between survival under the two regimes.

\begin{center}
    \textit{Results}
\end{center}
%\subsection{Main results}
We estimated the $PCIE$ and the $SAIE$ using L-DART with priors on the sensitivity parameters as described above. For each chain we used a thinning of 100, discarded the first 300 iterations as burn-in, and kept 2,400 posterior samples for inference. We sampled pseudo data of size $N^*=10,000$ at each iteration for the MC integration. Convergence of the posterior samples was monitored using trace plots. Note, the memory outcome, i.e. a test score, is presented in \%. 

Results for the different age-cohorts are shown in Figure 1. The results revealed that posterior mean episodic memory score at end of follow-up was attenuated with advancing age, both under the natural course and under the hypothetical intervention, (although to a lesser extent for the latter). Among subjects 70 and older, the shift in memory from the intervention corresponds to postponing memory decline approximately $1.5$ years on average, although the effect is non-significant. The results for the $PCIE$ suggest that there is no long-term cognitive effects of monitoring sBP in this population (credible intervals cover 0 for all age-cohorts), although the effect is notably stronger for the older age-cohorts compared to middle-aged. The estimates of the $PCIE$ and the $SAIE$ were similar. For the $SAIE$ this suggests that there is no long-term cognitive effects in the always survivor stratum, i.e. the stratum where individuals were to survive under both the natural course and the intervention. Hence, the non-significant results for the $PCIE$ are not driven by differential mortality rates. We moreover note that only slightly higher survival probabilities were obtained under the the hypothetical intervention for all age cohorts
when compared to survival probabilities under the natural course, and as a result, leading to rather small differences between the $PCIE$ and the $SAIE$. 

We performed sensitivity analyses to investigate robustness of results to violations of the assumptions C2-C6. Details are given in Appendix D and results are summarized in Table \ref{table_betula_SA}. To ease presentation we divided the sample into two age groups; middle-aged, i.e. $age_{iT}< 65$, and a older group, i.e $age_{iT}\geq 65$. 

First, the results are not sensitive to other (plausible) specifications of the priors on $\lambda$ and $\Delta$, data not shown. Moreover, the results were consistent with the results obtained using default assumptions for both age groups when taking into account a unmeasured time-varying confounder. We further investigated the impact of possible positivity violations on the results. For this analysis, 61 subjects with limited overlap were discarded. The results did not differ significantly from the main analysis, although the effect was somewhat attenuated for both middle aged and older. 
Using the missing not at random (MNARS) assumption instead of the MARS for the outcome resulted in attenuated posterior mean episodic memory score for both regimes. However, the estimates of the $PCIE$ were stable. 
Finally, the results did not differ significantly from the main analysis when fitting the two other \textit{soft} BART models, i.e. S-DART and S-DART-lag1, instead of LDART. However, the effects were much smaller for both middle aged and older. 

%%%%%%%%%%%%%%%%%%%%%%%%%%%%%%%%%%%%%%%%%%%%%%%%%%%%%%%%%%%%%

\begin{center}
    \textsc{8 Conclusions}
\end{center}
%\section{Conclusions}
Using data from a prospective cohort study on aging, cognition and dementia, we investigated population effects on episodic memory of an hypothetical incremental threshold intervention over 15 years that reduced sBP in individuals with hypertension.
In our study, there were no significant prognostic- or etiological effects, although effects were markedly stronger in older individuals. These findings provide important information about long-term blood pressure control in cognitive aging, indicating that sBP interventions likely do not have a strong effect on memory, whether at the overall population level or in the population that would survive under both the natural course and the intervention. However, more research from long-term clinical trials on whether lowering sBP can prevent cognitive decline is needed to draw firm conclusions.

%Previous clinical trials on sBP lowering interventions yielded mixed findings  (e.g. \citeauthor{sprint_mci} \citeyear{sprint_mci}; \citeauthor{hope3} \citeyear{hope3}). Our results align with studies that did not find a significant association, and possible explanations include variations in cognitive measures, follow-up duration, and study populations. However, our study, utilizing a conservative but more realistic definition of an intervention, may be less affected by baseline confounding compared to previous prospective studies (e.g. \cite{launer1995}).  Additionally, the potential benefit of more intensive sBP control (sBP $<120$ mm Hg; e.g. \cite{sprint_mci}), remains unexplored in the current study due to data limitations. %Hence, future interventions should consider more intensive sBP control.

We propose a Bayesian semi-parametric approach for the proposed incremental threshold intervention, extending prior research in several ways. First, our approach introduces an incremental intervention for prospective cohort data with attrition, offering a viable alternative when clinical trials are not available or feasible to conduct. Second, we offer two types of mortal cohort inferences, providing both prognostic and etiological effect estimates, both of which are critical for public health practice, especially in older populations. Last, our modeling approach employs \textit{soft} BART models with a novel sparsity-inducing hyperprior for longitudinal data, promoting parsimony in longitudinal prediction settings and demonstrating good performance in simulations.  

Concerns have been raised about principal stratification methods relying on assumptions that are often hard to justify in practice (e.g. \cite{feller2017principal}). Monotonicity is one such assumption. Our approach relies on the less strong \textit{stochastic monotonicity} assumption and uses sensitivity parameters that allow researcher to investigate sensitivity of results to this unverifiable assumption. However, a limitation with proposed principal stratification approach is that we only included age as a survival-predictive covariate for partial identification of the SAIE. In future work we will explore the possibility of incorporating more baseline covariates (see for example \cite{roy2008principal, tchetgen2014identification, wang2017deaths}).

\begin{center}
    \textsc{Acknowledgments}
\end{center}
%\section*{Acknowledgments}
This work is funded by Forte Dnr 2019-01064 to MJ, and NIH RO1 HL166324 and HL158963 to MJD. This research was conducted using the resources of High Performance Computing Center North (HPC2N). The Betula Project is supported by Knut and Alice Wallenberg foundation (KAW) and the Swedish Research Council (K2010-61X-21446-01).

\begin{center}
    \textsc{Supplementary material}
\end{center}
%\section*{Supplementary material}
The Supplementary Appendices A and B provide details for the extended G-formula and identiﬁcation of the SAIE. Computational details for the longitudinal Dirichlet prior and the G-computation algorithm are given in Appendix C. Supplementary Betula results are 
given in Appendix D.

\begin{center}
\textsc{References}
\end{center}
\printbibliography[heading=none]

\clearpage
\begin{center}
    \textsc{Tables and Figure}
\end{center}
%\section{Tables and Figure}

\begin{table}[H]
\small
\centering
\caption{Simulation results for estimating $E_{H^{g^{nc}}}\{E[Y_T(g^{nc}) \mid H^{g^{nc}}]\}$ using (\ref{gformula}) and the simulated data as described in Section 6 with 50 regression trees. For scenario A: $\bar{c}=(1,0.75,0.5,0.25)$, B: $\bar{c}=(1,0.75,0.5,0.25)^3$, C: $\bar{c}=(1,0,0,0)$, and D: $\bar{c}=(0.25,0.25,0.25,0.25)$. The results are presented in terms of absolute relative bias $\times 100$, root-mean-squared-error (rmse) and 95\% coverage probabilities (cp).} 
\begin{tabular}{l ccc ccc ccc ccc}
  \hline
   &  \multicolumn{3}{c}{A}  &  \multicolumn{3}{c}{B}  &  \multicolumn{3}{c}{C}  &  \multicolumn{3}{c}{D}  \\ 
  \hline
 & bias & rmse & cp & bias & rmse & cp & bias & rmse & cp & bias & rmse & cp \\ 
  \hline
   n=1500  &  &  &  &  &  &  &  &  &  & &  &  \\ 
   \hline
  LBART & 1.1 & 0.49 & 0.93 & 1.4 & 0.41 & 0.94 & 2.1 & 0.38 & 0.96 & 1.9 & 0.35 & 0.97 \\ 
  SBART & 1.0 & 0.45 & 0.96 & 1.4 & 0.42 & 0.95 & 2.1 & 0.39 & 0.95 & 2.0 & 0.35 & 0.97 \\ 
  SBART-lag1 & 1.1 & 0.49 & 0.93 & 1.4 & 0.39 & 0.94 & 2.1 & 0.38 & 0.96 & 2.0 & 0.37 & 0.94 \\ 
  HDART & 1.1 & 0.51 & 0.97 & 1.7 & 0.48 & 0.94 & 2.4 & 0.44 & 0.96 & 2.2 & 0.40 & 0.95 \\ 
  HBART & 1.2 & 0.52 & 0.97 & 1.7 & 0.47 & 0.97 & 2.3 & 0.43 & 0.97 & 2.3 & 0.42 & 0.98 \\ 
  \hline
   n=750  &  &  &  &  &  &  &  &  &  & &  &  \\ 
  \hline
  LBART & 1.3 & 0.62 & 0.95 & 1.9 & 0.55 & 0.96 & 2.9 & 0.52 & 0.96 & 2.6 & 0.49 & 0.95 \\ 
  SBART & 1.4 & 0.63 & 0.96 & 2.0 & 0.55 & 0.96 & 3.0 & 0.55 & 0.95 & 2.6 & 0.50 & 0.94 \\ 
  SBART-l1 & 1.4 & 0.65 & 0.95 & 1.9 & 0.55 & 0.94 & 3.0 & 0.55 & 0.93 & 2.8 & 0.51 & 0.92 \\ 
  HDART & 1.5 & 0.70 & 0.95 & 2.2 & 0.62 & 0.96 & 3.1 & 0.57 & 0.96 & 3.0 & 0.55 & 0.96 \\ 
  HBART & 1.6 & 0.73 & 0.94 & 2.2 & 0.62 & 0.96 & 3.4 & 0.61 & 0.97 & 3.1 & 0.55 & 0.97 \\  
  \hline
   n=250  &  &  &  &  &  &  &  &  &  & &  &  \\ 
  \hline
  LBART & 2.2 & 1.03 & 0.95 & 3.2 & 0.91 & 0.96 & 4.8 & 0.88 & 0.95 & 4.7 & 0.85 & 0.92 \\ 
  SBART & 2.4 & 1.07 & 0.96 & 3.3 & 0.93 & 0.95 & 4.9 & 0.89 & 0.94 & 4.7 & 0.84 & 0.93 \\ 
  SBART-l1 & 2.4 & 1.11 & 0.93 & 3.1 & 0.88 & 0.96 & 4.8 & 0.87 & 0.95 & 4.8 & 0.85 & 0.91 \\ 
  HDART & 2.5 & 1.12 & 0.95 & 3.5 & 1.00 & 0.95 & 5.2 & 0.95 & 0.96 & 4.8 & 0.87 & 0.96 \\ 
  HBART & 2.4 & 1.10 & 0.95 & 3.5 & 0.99 & 0.96 & 5.2 & 0.95 & 0.97 & 4.9 & 0.89 & 0.97 \\ 
  \hline
\end{tabular}
\label{table_sim_study_1}
\end{table}

% Figure 1
\begin{figure}[hbt!]
\includegraphics[width=1.0\textwidth]{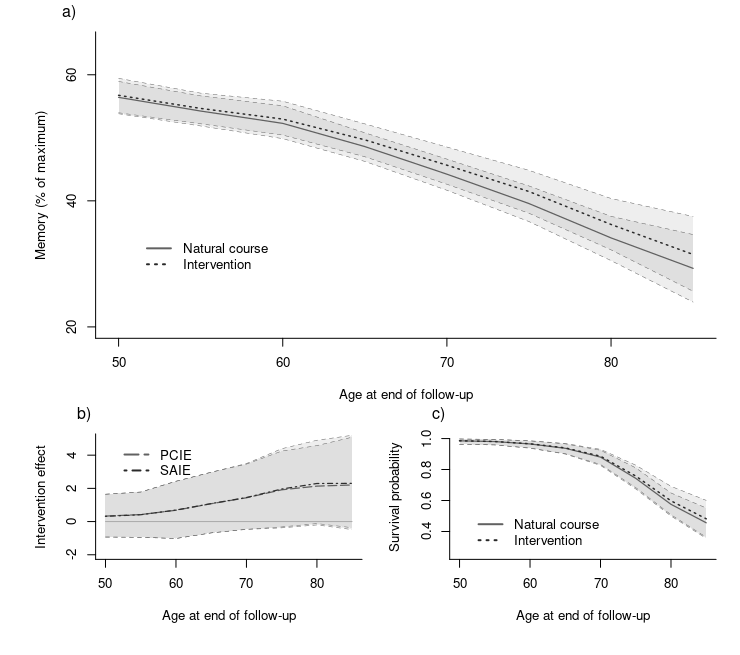} 
\caption{Results from the main analysis of Betula data using default assumptions and sensitivity parameters as described in Section 7. Panel a) Age-specific posterior mean memory at end of follow-up under the natural course (plain line) and under the hypothetical intervention (dotted line). Panel b) Results for the $PCIE$ and the $SAIE$. Panel c) Posterior mean survival probabilities at end of follow-up under the natural course and the hypothetical intervention. All plots include 95\% pointwise credible intervals.} 
\label{Fig_Betula} 
\end{figure}

\clearpage
\thispagestyle{plain}
\begin{center}
    {\Large
    \textbf{Supplementary material - Long-term cognitive effects of an incremental blood pressure intervention in a mortal cohort}}
        
\vspace{0.5cm}
\textsc{Josefsson, Maria, Karalija, Nina, \& Daniels, Michael J.}
\end{center}
\vspace{0.5cm}

\begin{center}
    \textsc{A. The Extended G-formula}
\end{center}
%\subsection*{1. The Extended G-formula}
Below we provide details of the Extended G-formula as introduced in Section 4.2.
\begin{align}\label{extgformula}
%& E_{H_{int^*}}\{E[Y_T | S_T=1, H_{int^*}]\} = \nonumber \\
& \int_{\bar{s}_{t}}\int_{\bar{r}_{t}}\int_{\bar{l}_{t}}\int_{\bar{a}_{t}^*}\int_{\bar{a}_{t}}\int_{\bar{y}_{t-1}} E[Y_T  | \bar{y}_{T-1}, \bar{a}_{T}, \bar{l}_T, \bar{r}_{T}, \bar{s}_{T}=1] \nonumber \\
& \quad \times \prod_{k=1}^{T} p^{int^*}(a_k | \bar{y}_{k-1}, a^*_{k}, \bar{a}_{k-1}, \bar{l}_{k}, \bar{s}_{k}=1) \nonumber \\ 
& \quad \times p(a^*_k | \bar{y}_{k-1}, d_{k-1}, \bar{l}_{k}, \bar{s}_{k}=1) \nonumber \\ 
& \quad \times p(l_k | \bar{y}_{k-1}, \bar{a}_{k-1}, \bar{l}_{k-1}, \bar{r}_{k}, \bar{s}_{k}=1) \nonumber \\ 
& \quad \times p(r_k | \bar{y}_{k-1}, \bar{a}_{k-1}, \bar{l}_{k-1}, \bar{r}_{k-1}, \bar{s}_{k}=1) \nonumber \\ 
& \quad \times p(s_k | \bar{y}_{k-1}, \bar{a}_{k-1}, \bar{l}_{k-1}, \bar{r}_{k-1}, \bar{s}_{k-1}=1) \nonumber \\ 
& \quad \times p(y_{k-1} | \bar{y}_{k-2}, \bar{a}_{k-1}, \bar{l}_{k-1}, \bar{r}_{k-1}, \bar{s}_{k-1}=1). \tag{A.1} %\nonumber \\
%& \qquad d\bar{y}_{t-1} d\bar{a}_t d\bar{l}_t d\bar{r}_t d\bar{s}_t.
\end{align}

\begin{center}
    \textsc{B. Identification of the SAIE under a hypothetical intervention based on the natural value of treatment}
\end{center}
%\subsection*{A. Identification of the SAIE under a hypothetical intervention based on the natural value of treatment}

%%%%%%%%%%%%%%%%%%%%%
Identification of the SAIE under a hypothetical intervention based on the NVT under a stochastic Monotonicity assumption is presented below. 
First, for the two contrasting regimes $g^{nc}$ and $g^{int^*}$, we have that the expected potential outcome for the principal strata becomes,
\begin{align*}\label{sace.eq2}
&E[Y_T(g^{int^*}) - Y_T(g^{nc}) \mid \bar{S}(g^{nc})=S_T(g^{int^*})=1] = \nonumber \\
& \quad PCIE + \Delta \left\{ \psi^{g^{int^*}} + \lambda \left(U - \psi^{g^{int^*}}\right) \right\} \left(1 - U^{-1} \right), \tag{A.2}
\end{align*}
where $U=\min\left\{1, \frac{\psi^{g^{int^*}}}{\psi^{g^{nc}}}\right\}$.
Identification of [\ref{sace.eq2}] is obtained by using the law of total probability (ltp) and some algebra. First, for the intervention regime $g^{int^*}$, and similarly for $g^{nc}$, we have that
\begin{align*}
E[&Y_T(g^{int^*})) \mid S_T(g^{int^*})=1] = \\
 & \quad E[Y_T(g^{int^*}) \mid S_T(g^{nc})=S_T(g^{int^*})=1] + \Pr[S_T(g^{nc})\neq 1 \mid S_T(g^{int^*})=1] \times \\
& \quad \left\{E[Y_T(g^{int^*}) \mid S_T(g^{nc})=1, S_T(g^{int^*})\neq 1] - E[Y_T(g^{int^*}) \mid \bar{S}(g^{nc})=S_T(g^{int^*})=1]\right\}. 
\end{align*}
Using A6 and solving the above equation for $E[Y_T(g^{int^*}) \mid S_T(g^{nc})=S_T(g^{int^*})=1]$ we have that 
\begin{align*}\label{Solv2}
E[&Y_T(g^{int^*}) \mid \bar{S}(g^{nc})=S_T(g^{int^*})=1] = \\
&E[Y_T(g^{int^*}) \mid S_T(g^{int^*})=1] + \Delta \times \left\{1 - \Pr[S_T(g^{int^*}) = 1 \mid S_T(g^{nc})=1] \right\},
\end{align*}
where $\Delta$ is a sensitivity parameter when comparing the difference in potential outcomes when comparing the principal strata to the strata where subjects were to live under the hypothetical intervention but not under the natural course. Hence, 
\begin{align*}
    &E[Y_T(g^{int^*}) - Y_T(g^{nc}) \mid \bar{S}(g^{nc})=S_T(g^{int^*})=1] \\
    & \quad = E[Y_T(g^{int^*}) \mid S_T(g^{int^*})=1] + \Delta \times \left\{1-\Pr[S_T(g^{nc})= 1 \mid S_T(g^{int^*})=1]\right\} \\
    & \qquad - E[Y_T(g^{nc}) \mid S_T(g^{nc})=1] - \Delta \times \left\{1-\Pr[S_T(g^{int^*})= 1 \mid S_T(g^{nc})=1]\right\} \\
    & \quad = E[Y_T(g^{int^*}) \mid S_T(g^{int^*})=1] - E[Y_T(g^{nc}) \mid S_T(g^{nc})=1] + \\
    & \qquad \Delta \times \left\{\Pr[S_T(g^{int^*})= 1 \mid S_T(g^{nc})=1]-\Pr[S_T(g^{nc})= 1 \mid S_T(g^{int^*})=1]\right\} \\
    & \quad = PCIE + \Delta \times \left\{\Pr[S_T(g^{int^*})= 1 \mid S_T(g^{nc})=1]-\Pr[S_T(g^{nc})= 1 \mid S_T(g^{int^*})=1]\right\}
\end{align*}
Moreover, by using A5 (Stochastic monotonicity) and Bayes theorem we have that \\
$\Pr[S_T(g^{int^*}) = 1 \mid  S_T(g^{nc}) = 1] =  \psi^{g^{int^*}} + \lambda \left(U - \psi^{g^{int^*}}\right)$ 
and
$\Pr[S_T(g^{nc}) = 1 \mid  S_T(g^{int^*}) = 1] = \frac{\psi^{g^{nc}}}{\psi^{g^{int^*}}} \left[\psi^{g^{int^*}} + \lambda \left(U - \psi^{g^{int^*}}\right)\right]$,
where $\lambda$ is a sensitivity parameter, and $0\leq \lambda \leq 1$. Hence, the SAIE simplifies to (A.2).

%%%%
For $g\in(g^{int^*}, g^{nc})$ and using A1-A4, we have that
\begin{align}
E[Y_T(g) \mid S_T(g)=1]& = \frac{E[Y(g), S_T(g)=1]}{\Pr[S_T(g)=1]} \nonumber \\
& = \frac{E_{H^{g}}[E(Y_T, S_T=1\mid H^{g})]}{E_{H^{g}}[\Pr(S_T=1\mid H^{g})]}.  \tag{A.3}
\end{align}
Under the natural course (${g^{nc}}$), $E_{H^{g^{nc}}}[E(Y_T, S_T=1\mid H^{g^{nc}})]$ is obtained by marginalizing over the  distributions of the set of temporally preceding variables $H^{g^{nc}}$ using Robins (\citeyear{robins1986new}) g-formula. Thus, for $g^{nc}$ we have that
\begin{align*}
E&[Y_T(g^{nc}), S_T(g^{nc})=1]=E_{H^{g^{nc}}}\{E[Y_T(g^{nc}), S_T(g^{nc})=1 \mid H^{g^{nc}}]\} \\
& = \int \ldots \int E[Y_T(g^{nc}) \mid \bar{S}_T=1, \bar{Y}_{T-1}=\bar{y}_{T-1}, \bar{L}_T=\bar{l}_T, \bar{R}_{T}=\bar{r}_{T}] \\
& \quad \times \Pr[\bar{S}_T(g^{nc})=1 \mid \bar{Y}_{T-1}=\bar{y}_{T-1}, \bar{L}_{T-1}=\bar{l}_{T-1}, \bar{R}_{T-1}=\bar{r}_{T-1}] \\ 
& \quad \times \prod_{k=1}^{T} p(Y_{k-1}(g^{nc})=y_{k-1} \mid \bar{Y}_{k-2}=\bar{y}_{k-2}, \bar{L}_{k-1}=\bar{l}_{k-1}, \bar{R}_{k-1}=\bar{r}_{k-1}, \bar{S}_{k-1}=1) \\
& \qquad \times p(A_k(g^{nc})=a_k \mid \bar{Y}_{k-1}=\bar{y}_{k-1}, \bar{L}_{k}=\bar{l}_{k}, \bar{R}_{k}=\bar{r}_{k},  \bar{S}_{k}=1) \\ 
& \qquad \times p(L_k(g^{nc})=l_k \mid \bar{Y}_{k-1}=\bar{y}_{k-1}, \bar{L}_{k-1}=\bar{l}_{k-1}, \bar{R}_{k-1}=\bar{r}_{k-1}, \bar{S}_{k}=1) \\ 
& \qquad \times p(R_k(g^{nc})=r_k \mid \bar{Y}_{k-1}=\bar{y}_{k-1}, \bar{L}_{k-1}=\bar{l}_{k-1}, \bar{R}_{k-1}=\bar{r}_{k-1}, \bar{S}_{k}=1) d\bar{y}_{T-1} d\bar{a}_T d\bar{l}_T d\bar{r}_T \\
& \quad \text{Exchangeability} \\ 
& = \int \ldots \int E[Y_j(g^{nc}) \mid \bar{S}_{t}=1, \bar{Y}_{t-1}=\bar{y}_{t-1}, \bar{A}_{t}=z_{t}, \bar{L}_j=\bar{l}_j, \bar{R}_{T}=\bar{r}_{T}] \\
& \quad \times \Pr[\bar{S}_{t}(g^{nc})=1 \mid \bar{Y}_{t-1}=\bar{y}_{t-1}, \bar{A}_{t-1}=z_{t-1}, \bar{L}_{t-1}=\bar{l}_{t-1}, \bar{R}_{T-1}=\bar{r}_{T-1}] \\ 
& \quad \times \prod_{k=1}^T p(y_{k-1} \mid \bar{Y}_{k-2}=\bar{y}_{k-2}, \bar{A}_{k-1}=z_{k-1}, \bar{L}_{k-1}=\bar{l}_{k-1}, \bar{R}_{k-1}=\bar{r}_{k-1}, \bar{S}_{k-1}=1) \\
& \qquad \times p(a_k \mid \bar{Y}_{k-1}=\bar{y}_{k-1}, \bar{A}_{k-1}=z_{k-1}, \bar{L}_{k}=\bar{l}_{k}, \bar{R}_{k}=\bar{r}_{k},  \bar{S}_{k}=1) \\ 
& \qquad \times p(l_k \mid \bar{Y}_{k-1}=\bar{y}_{k-1}, \bar{A}_{k-1}=z_{k-1}, \bar{L}_{k-1}=\bar{l}_{k-1}, \bar{R}_{k-1}=\bar{r}_{k-1}, \bar{S}_{k}=1) \\ 
& \qquad \times p(r_k \mid \bar{Y}_{k-1}=\bar{y}_{k-1}, \bar{A}_{k-1}=\bar{a}_{k-1}, \bar{L}_{k-1}=\bar{l}_{k-1}, \bar{R}_{k-1}=\bar{r}_{k-1}, \bar{S}_{k}=1) d\bar{y}_{T-1} d\bar{a}_T d\bar{l}_T d\bar{r}_T \\
& \quad \text{Consistency} \\ 
& = \int \ldots \int E[Y_j \mid \bar{S}_{t}=1, \bar{Y}_{t-1}=\bar{y}_{t-1}, \bar{A}_{t}=z_{t}, \bar{L}_j=\bar{l}_j, \bar{R}_{T}=\bar{r}_{T}] \\
& \quad \times \Pr[\bar{S}_{t}=1 \mid \bar{Y}_{t-1}=\bar{y}_{t-1}, \bar{A}_{t-1}=z_{t-1}, \bar{L}_{t-1}=\bar{l}_{t-1}, \bar{R}_{T-1}=\bar{r}_{T-1}] \\ 
& \quad \times \prod_{k=1}^T p(y_{k-1} \mid \bar{Y}_{k-1}=\bar{y}_{k-1}, \bar{A}_{k-1}=z_{k-1}, \bar{L}_{k-1}=\bar{l}_{k-1}, \bar{R}_{k-1}=\bar{r}_{k-1}, \bar{S}_{k-1}=1) \\
& \qquad \times p(a_k \mid \bar{Y}_{k-1}=\bar{y}_{k-1}, \bar{A}_{k-1}=z_{k-1}, \bar{L}_{k}=\bar{l}_{k}, \bar{R}_{k}=\bar{r}_{k},  \bar{S}_{k}=1) \\ 
& \qquad \times p(l_k \mid \bar{Y}_{k-1}=\bar{y}_{k-1}, \bar{A}_{k-1}=z_{k-1}, \bar{L}_{k-1}=\bar{l}_{k-1}, \bar{R}_{k-1}=\bar{r}_{k-1}, \bar{S}_{k}=1) \\ 
& \qquad \times p(r_k \mid \bar{Y}_{k-1}=\bar{y}_{k-1}, \bar{A}_{k-1}=\bar{a}_{k-1}, \bar{L}_{k-1}=\bar{l}_{k-1}, \bar{R}_{k-1}=\bar{r}_{k-1}, \bar{S}_{k}=1) d\bar{y}_{T-1} d\bar{a}_T d\bar{l}_T d\bar{r}_T  \\
& = E_{H^{g^{nc}}}\{E[Y, S_T=1\mid H^{g^{nc}}]\}=E[Y, S_T=1].
\end{align*}
A similar argument shows that $\Pr[S_T(g^{nc})=1] = E_{H_{int}}\{ \Pr[S_T(g^{nc})=1\mid H_{int}]\} = E_{H_{int}}\{ \Pr[S_T=1\mid H_{int}]\}=\Pr[S_T=1]$, and therefore,
\begin{align}
\frac{E[Y(g^{nc}), S_T(g^{nc})=1]}{\Pr[S_T(g^{nc})=1]} = E_{g^{nc}}[Y \mid S_T=1].  \tag{A.4}
\end{align}
Hence, it follows that (A.4) can be estimated using the g-formula conditioning on survival at a specific time point as described in (\ref{gformula}) in the main text.

%%%%%%%%%%%%%%%%%%%%%%%%%%%%%%%%%%%%%%%%%
%%%%%%%%%%%%%%%%%%%%%%%%%%%%%%%%%%%%%%%%%
For regime $g^{int^*}$, $E_{H^{g^{int^*}}}[E(Y_T, S_T=1\mid H^{g^{int^*}})]$ is also obtained by marginalizing over the  distributions of the set of temporally preceding variables $H^{g^{int^*}}$. However, the extended g-formula must be used for a intervention that depends on the NVT \parencite{robins2004}. As such, for regime $g^{int^*}$ we have that
%\newpage\vspace*{-3.75cm}
\begin{align*}
E&(Y_T(g^{int^*}), S_T(g^{int^*})=1)=E_{H^{g^{int^*}}}[E(Y_T(g^{int^*}), S_T(g^{int^*})=1)\mid H^{g^{int^*}}] \\
& = \int \ldots \int E[Y_T(g^{int^*}) \mid \bar{S}_T=1, \bar{Y}_{T-1}=\bar{y}_{T-1}, \bar{L}_T=\bar{l}_T, \bar{R}_{T}=\bar{r}_{T}] \\
& \quad \times \Pr[\bar{S}_T(g^{int^*})=1 \mid \bar{Y}_{T-1}=\bar{y}_{T-1}, \bar{L}_{T-1}=\bar{l}_{T-1}, \bar{R}_{T-1}=\bar{r}_{T-1}] \\ 
& \quad \times \prod_{k=1}^{T} p(Y_{k-1}(g^{int^*})=y_{k-1} \mid \bar{Y}_{k-2}=\bar{y}_{k-2}, \bar{L}_{k-1}=\bar{l}_{k-1}, \bar{R}_{k-1}=\bar{r}_{k-1}, \bar{S}_{k-1}=1) \\
& \qquad \times p^{int^*}(a_k \mid \bar{Y}_{k-1}=\bar{y}_{k-1}, A^*_{k}=a^*_{k}, \bar{A}_{k-1}=\bar{a}_{k-1}, \bar{L}_{k}=\bar{l}_{k}, \bar{R}_{k}=\bar{r}_{k},  \bar{S}_{k}=1) \\ 
& \qquad \times p(A^*_k(g^{int^*})=a^*_k \mid \bar{Y}_{k-1}=\bar{y}_{k-1}, \bar{L}_{k}=\bar{l}_{k}, \bar{R}_{k}=\bar{r}_{k},  \bar{S}_{k}=1) \\ 
& \qquad \times p(L_k(g^{int^*})=l_k \mid \bar{Y}_{k-1}=\bar{y}_{k-1}, \bar{L}_{k-1}=\bar{l}_{k-1}, \bar{R}_{k-1}=\bar{r}_{k-1}, \bar{S}_{k}=1) \\ 
& \qquad \times p(R_k(g^{int^*})=r_k \mid \bar{Y}_{k-1}=\bar{y}_{k-1}, \bar{L}_{k-1}=\bar{l}_{k-1}, \bar{R}_{k-1}=\bar{r}_{k-1}, \bar{S}_{k}=1) d\bar{y}_{T-1} d\bar{a}^*_T d\bar{a}_T d\bar{l}_T d\bar{r}_T \\
& = \ldots
\end{align*}
\newpage\vspace*{-0.0cm}
\begin{align*}
& \ldots \\
& \quad \text{Exchangeability} \\ 
& = \int \ldots \int E[Y_T(g^{int^*}) \mid \bar{S}_{T}=1, \bar{Y}_{T-1}=\bar{y}_{T-1}, \bar{A}_{T}=z_{T}, \bar{L}_T=\bar{l}_T] \\
& \quad \times \Pr[\bar{S}_{T}(g^{int^*})=1 \mid \bar{Y}_{T-1}=\bar{y}_{T-1}, \bar{A}_{T-1}=a_{T-1}, \bar{L}_{T-1}=\bar{l}_{T-1}, \bar{R}_{T-1}=\bar{r}_{T-1}] \\ 
& \quad \times \prod_{k=1}^T p(y_{k-1} \mid \bar{Y}_{k-2}=\bar{y}_{k-2}, \bar{A}_{k-1}=z_{k-1}, \bar{L}_{k-1}=\bar{l}_{k-1}, \bar{R}_{k-1}=\bar{r}_{k-1}, \bar{S}_{k-1}=1) \\
& \qquad \times p^{int^*}(a_k \mid \bar{Y}_{k-1}=\bar{y}_{k-1}, A^*_{k}=a^*_{k}, \bar{A}_{k-1}=z_{k-1}, \bar{L}_{k}=\bar{l}_{k}, \bar{R}_{k}=\bar{r}_{k},  \bar{S}_{k}=1) \\ 
& \qquad \times p(a^*_k \mid \bar{Y}_{k-1}=\bar{y}_{k-1}, \bar{A}_{k-1}=z_{k-1}, \bar{L}_{k}=\bar{l}_{k}, \bar{R}_{k}=\bar{r}_{k},  \bar{S}_{k}=1) \\ 
& \qquad \times p(l_k \mid \bar{Y}_{k-1}=\bar{y}_{k-1}, \bar{A}_{k-1}=z_{k-1}, \bar{L}_{k-1}=\bar{l}_{k-1}, \bar{R}_{k-1}=\bar{r}_{k-1}, \bar{S}_{k}=1) \\ 
& \qquad \times p(r_k \mid \bar{Y}_{k-1}=\bar{y}_{k-1}, \bar{A}_{k-1}=\bar{a}_{k-1}, \bar{L}_{k-1}=\bar{l}_{k-1}, \bar{R}_{k-1}=\bar{r}_{k-1}, \bar{S}_{k}=1) d\bar{y}_{T-1} d\bar{a}^*_T d\bar{a}_T d\bar{l}_T d\bar{r}_T \\
& \quad \text{Consistency} \\ 
& = \int \ldots \int E[Y_T \mid \bar{S}_{T}=1, \bar{Y}_{T-1}=\bar{y}_{T-1}, \bar{A}_{T}=a_{T}, \bar{L}_T=\bar{l}_T, \bar{R}_{T}=\bar{r}_{T}] \\
& \quad \times \Pr[\bar{S}_{T}=1 \mid \bar{Y}_{T-1}=\bar{y}_{T-1}, \bar{A}_{T-1}=a_{T-1}, \bar{L}_{T-1}=\bar{l}_{T-1}, \bar{R}_{T-1}=\bar{r}_{T-1}] \\ 
& \quad \times \prod_{k=1}^T p(y_{k-1} \mid \bar{Y}_{k-1}=\bar{y}_{k-1}, \bar{A}_{k-1}=z_{k-1}, \bar{L}_{k-1}=\bar{l}_{k-1}, \bar{R}_{k-1}=\bar{r}_{k-1}, \bar{S}_{k-1}=1) \\
& \qquad \times p^{int^*}(a_k \mid \bar{Y}_{k-1}=\bar{y}_{k-1}, A^*_{k}=a^*_{k}, \bar{A}_{k-1}=\bar{a}_{k-1}, \bar{L}_{k}=\bar{l}_{k}, \bar{R}_{k}=\bar{r}_{k},  \bar{S}_{k}=1) \\ 
& \qquad \times p(a^*_k \mid \bar{Y}_{k-1}=\bar{y}_{k-1}, \bar{A}_{k-1}=z_{k-1}, \bar{L}_{k}=\bar{l}_{k}, \bar{R}_{k}=\bar{r}_{k},  \bar{S}_{k}=1) \\ 
& \qquad \times p(l_k \mid \bar{Y}_{k-1}=\bar{y}_{k-1}, \bar{A}_{k-1}=z_{k-1}, \bar{L}_{k-1}=\bar{l}_{k-1}, \bar{R}_{k-1}=\bar{r}_{k-1}, \bar{S}_{k}=1) \\ 
& \qquad \times p(r_k \mid \bar{Y}_{k-1}=\bar{y}_{k-1}, \bar{A}_{k-1}=\bar{a}_{k-1}, \bar{L}_{k-1}=\bar{l}_{k-1}, \bar{R}_{k-1}=\bar{r}_{k-1}, \bar{S}_{k}=1) d\bar{y}_{T-1} d\bar{a}^*_T d\bar{a}_T d\bar{l}_T d\bar{r}_T  \\
& = E_{H^{g^{int^*}}}[E(Y_T, S_T=1\mid H^{g^{int^*}})]=E(Y_T, S_T=1).
\end{align*}
%%%%%%%%%%%%%%%%%%%%%%%%%%%%%%%%%%%%%%%%%
%%%%%%%%%%%%%%%%%%%%%%%%%%%%%%%%%%%%%%%%%
Similarly as for $g^{nc}$ we have that $\Pr[S_T(g^{int^*})=1] = E_{H^{g^{int^*}}}\{ \Pr[S_T(g^{int^*})=1\mid H^{g^{int^*}}]\}=E_{H^{g^{int^*}}}\{ \Pr[S_T=1\mid H^{g^{int^*}}]\}=\Pr[S_T=1]$, and therefore,
\begin{align}\label{expYS22}
\frac{E[Y(g^{int^*}), S_T(g^{int^*})=1]}{\Pr[S_T(g^{int^*})=1]} = E_{g^{int^*}}[Y \mid S_T=1].  \tag{A.5}
\end{align}
Hence, it follows that (A.5) can be estimated using the extended g-formula conditioning on survival at a specific time point as described in (\ref{extgformula}) in the main text. Moreover, by combining the results above we show that the SAIE become (A.2).

\begin{center}
    \textsc{C. Computational details}
\end{center}
%\section*{B. Computational details}
%\subsubsection*{Posterior computations}
\begin{center}
    \textit{Posterior computations}
\end{center}
Similar to Linero and Yang (2018), we obtain simple conjugate posterior updates for $(w, v^t, u, v^k)$ by implementing the Bayesian backfitting approach to sample from the posterior using the Markov chain Monte Carlo algorithm \parencite{chipman2010bart}. In particular, the full conditional for $w$ is given by $w \sim Beta\left(a + m_{w_1}, b + m_{w_2}\right)$, where $m_{w_1}$ is the number of branch splits for the set of current predictors and $m_{w_2}$ is the number of branch splits for all the sets of preceding time-varying covariates.
Similarly, the full conditional of $\mathbf{v}^t$ and $\mathbf{v}^k$ are given by $\mathbf{v}^t \sim \mathcal{D}\left( \eta/P_t + m_{v^t_1}, \ldots, \eta/P_t  + m_{v^t_{P_t}} \right)$ and $\mathbf{v}^k \sim \mathcal{D}\left( \eta/P_k + m_{v^k_1}, \ldots, \eta/P_k  + m_{v^k_{P_k}} \right)$ respectively, where $m_{v^k_{j}}$ is the number of branch splits on the $j$th predictor in the set of predictors associated with time point $k=1,\ldots,t-1$.
Finally, the full conditional of $\mathbf{u}$ is given by $\mathbf{u} \sim D\left(\frac{\alpha_1}{t-1} + m_{u_1} ,\cdots,\frac{\alpha_{t-1}}{t-1} + m_{u_{t-1}}\right)$, where $m_{u_j}$ is the number of branch splits on the predictors in the $j$th set of time-varying covariates.

\newpage
\begin{center}
    \textit{G-computation algorithm}
\end{center}
%\subsection*{G-computation algorithm}
\begin{itemize}
\item[1.] \textit{Modeling the observed data}: For $t=0,\ldots,T$, sample from the observed data posteriors for the parameters of the conditional distributions of $y_t$, $a_t$, $l_t$, $r_t$ and $s_t$ using L-DART. Note, we implement the Bayesian bootstrap \parencite{rubin1981bayesianbootstrap} to integrate over the distribution of baseline confounders without missingness.
%%%%%%%%%%%%%%%%%%%%%%%%%%%%%%%%%%%%%%%%%%%%%%%%%%%%%%%%%%
\item[2.] \textit{Pseudo data}: For each posterior sample of the parameters of the L-DART models in Step 1, sequentially sample pseudo data, $D^{nc}$ and $D^{int^*}$, of size $N^p$ for the two contrasting regimes. %For the natural course this involves sequentially sampling: $D^{nc}=(l^{nc}_{0},a^{nc}_{0},y^{nc}_{0}, s_{1}^{nc}, r_{1}^{nc},\ldots, s_{T}^{nc}, r_{T}^{nc},l^{nc}_{T},a^{nc}_{T})$ for all $\bar{s}_{T}^{nc}=1$. 
Under the hypothetical intervention, to obtain pseudo data for $a_{t}$, pseudo NVT data is first sampled from the observed posterior distribution $a^{*}_{t}\sim N\left(\mu_{a_{t}}, \sigma_t^2\right)$ to compute $a_{t}= a^{*}_{t} + \delta_t I\{a^{*int}_{t}\geq \tau\}$, where a set of the shift parameter $\delta_t$ is first sampled from the associated prior distribution.
%
%\item[3.] \textit{MNAR missingness}: For both regimes, to allow for MNARS missingness in the outcome at time $t=2,\ldots,T$, we first sample one set from the prior distribution of the sensitivity parameter $\gamma_t$ and compute $y^{\boldsymbol{\cdot}}_t= \hat{y}^{\boldsymbol{\cdot}}_t - \gamma_tI\{r^{\boldsymbol{\cdot}}_t=0\}$ for all $\bar{s}_{T}^{\boldsymbol{\cdot}}=1$. 
%
\item[3.] \textit{G-computation}: Implement G-computation, as described in Section 3 for the two contrasting regimes using the pseudo data, $D^{nv}$ and $D^{int*}$, and sensitivity parameters, %$\bar{\gamma}_{t}$ and 
$\bar{\delta}_{t}$ from Step 2 to obtain posterior estimates, $\hat{\mu}^{int^*}$ and $\hat{\mu}^{nc}$, of $E_{H_{int^*}}\{E[Y_T | S_T=1, H_{int^*}]\}$ and $E_{H_{nc}}\{E[Y_T | S_T=1, H_{nc}]\}$ respectively. 
For the SAIE this also involves implementing G-computation for computing predicted survival at time $T$, i.e. $\hat{\psi}^{g}=\prod_{t=2}^T \hat{s}^{g}_{t}$ for the two contrasting regimes $g\in(g^{int^*}, g^{nc})$. 
\item[4.] \textit{Computation of the PCIE and the SAIE}: Implement Monte Carlo integration using the pseudo data for the two regimes to compute one posterior sample of the 
$$PCIE = \frac{\sum_i \hat{\mu}^{int^*}_{iT} s^{int^*}_{iT}}{\sum_t^T \sum_i s^{int^*}_{iT}} - \frac{\sum_i \hat{\mu}^{nc}_{iT} s^{nc}_{iT}}{\sum_t^T \sum_i s^{nc}_{iT}}$$ 
and the
$$SAIE = PCIE + \Delta \left\{ \hat{\psi}^{g^{int^*}} + \lambda \left(U - \hat{\psi}^{g^{int^*}}\right) \right\} \left(1 - U^{-1} \right),$$
where $U=\min\left\{1, \frac{\hat{\psi}^{g^{int^*}}}{\hat{\psi}^{g^{nc}}}\right\}$, one set of the sensitivity parameters $\Delta$ and $\lambda$ are samples.
\item[5.] Repeat step 2 - 4 for each posterior sample.
\end{itemize}

%\section*{C. Simulation results}

%\section*{D. Betula results}
\begin{center}
    \textsc{D. Supplementary Betula results}
\end{center}
\begin{table}[H]
\renewcommand\thetable{A1}
\centering
\caption{Baseline demographics for the middle-aged and old group in the Betula study, presented as mean (standard deviation) for the continuous variables and \% (N) for the dichotomous variables.}
\begin{tabular}{lcc}
  \hline
 & Middle aged & Older \\ 
  \hline
  Age & 42.5 (5.6) & 62.5 (5.6) \\ 
  Male & 0.46 (184) & 0.50 (198) \\ 
  Education & 12.6 (3.8) & 8.5 (3.2) \\ 
  sBP & 124.8 (15.6) & 147.7 (22.0) \\ 
  Smoker & 0.57 (230) & 0.53 (212) \\ 
  BMI & 24.3 (3.3) & 25.7 (3.5) \\ 
  Cholesterol & 52.0 (19.2) & 60.8 (19.3) \\ 
  Diabetes type 2 & 0.01 (6) & 0.07 (28) \\ 
  Episodic memory (\% correct) & 0.53 (0.12) & 0.42 (0.11) \\ 
   \hline
\end{tabular}
\label{betula_demographics}
\end{table}

\begin{center}
    \textit{Sensitivity analyses}
\end{center}
%\subsection{Sensitivity analyses}
Below we provide details of the sensitivity analyses to investigate robustness of results to violations of the assumptions C2-C4 and other BART specifications.

\textit{Unmeasured confounding:} We implement a simulation procedure to account for an unmeasured time-varying confounder missing for time-points $k > l$. In the Betula study, measures for cholesterol are obtained only for the first two time points; hence, it is a possible unmeasured confounder for $k > 2$. In the first step, $U_k$, for $k > 2$, is drawn from an estimate of the conditional distribution of $p(u_k \mid H^g_{u_k})$. Here, we assume $p(u_k \mid H^g_{u_k})=p(u_2 \mid H^g_{u_2})$, where $U_2$ is the last observed time point for $U_k$ and $H^g_{u_2}=(\bar{l}_2, \bar{r}_{2}, a_1, y_{1})$ is the history for $U_2$. 
To obtain a plausible fit for the conditional distribution of $p(u_k \mid H^g_{u_k})$, this involves first obtaining a fit for $p(u_2 \mid H^g_{u_2})$ using L-DART, then plugging $H^g_{u_k}$ into the L-DART model (for $u_2$), and, finally, drawing realizations for $U_k$ from the corresponding conditional distribution. Once we obtain realizations of $U_k$ for $k > 2$, we then fit the (extended) g-formula to the full data, and estimate the intervention effect as described in Section 4. Since $H^g_{u_2}$ and $H^g_{u_k}$ corresponds to different number of predictors, we incorporate only a subset of the predictors for $H^g_{u_k}$. In particular, we include $(r_{k}, l_{k}, y_{k-1}, a_{k-1}, l_{k-1}, r_{k-1})$ as well as baseline covariates. Note, $u_{k-1}$ is a subset of $l_{k-1}$ and is observed for $k=3$ and must first be estimated for $k>3$.
Since a single estimate corresponds to just one realization from the distribution of $U_k$, we simulate multiple draws (say 5) of $U_k$ for $k > 2$, and an overall estimate of the intervention effect is obtained by averaging over the single estimates.

\textit{Positivity violations:} To assess robustness of results to positivity violations (C3) we discard subjects with insufficient overlap. To identify these subjects we implemented a generalized propensity score approach, where the propensity score models at times $t=2,\ldots,T$ were fitted using L-DART. Subjects whose sBP was above the given threshold were excluded if their generalized propensity score were out of the range of the generalized propensity score among subjects with sBP within optimal levels. 

\textit{Missingness for the outcome:} Previous studies indicate that the dropout may be MNAR for the outcome \parencite{josefsson12}. As such, we introduce a sensitivity parameter $\gamma_{t}$ to identify the distribution for dropouts among survivors, to allow for deviations from the MARS assumption (C4). We assume $p(y_t | \bar{a}_{t}, \bar{l}_{t}, r_t=0, \bar{s}_{t}=1, \bar{y}_{t-1}) = p(y_t - \gamma_{t} | \bar{a}_{t}, \bar{l}_{t}, \bar{r}_{t}=1, \bar{s}_t=1, \bar{y}_{t-1})$ for all $t>0$. %If $\gamma_{t}=0$ this implies MARS on the distribution for the outcome. 
Our prior belief is that $\gamma_t<0$, reflecting a negative shift in memory score after a subject drops out. Here, the prior is specified as $\gamma_t \sim Triangular(L_{\gamma_t}, 0, L_{\gamma_t}),$ where we assume the lower bound not to be bigger than the estimated difference in memory change from baseline to follow-up between dropouts and completers, when adjusting for age and sex. That is, $\hat{E}[EMS_{i2}-EMS_{i1} | age_{i1}, sex_{i1}, r_{i3}=1] - \hat{E}[EMS_{i2}-EMS_{i1} | age_{i1}, sex_{i1}, r_{i3}=0]$, where $EMS_{it}$ is the episodic memory score at time $t$.

\textit{Other \textit{soft} BART specifications:} We compared the performance of our proposed approach (LDART) to the two \textit{soft} BART models; the standard SDART algorithm (as implemented in the R package SoftBart) and SoftBart incorporating only the predictors from the last time point in the model (SDART-lag1).

%\begin{center}
%\textit{Results of the sensitivity analyses}
%\end{center}
%\subsection{Results of the sensitivity analyses}
%To ease presentation we divided the sample into two age groups; middle-aged, i.e. $age_{iT}< 65$, and a older group, i.e $age_{iT}\geq 65$. The results are summarized in Table \ref{table_betula_SA}. 
%First, the results were consistent with the results obtained using default assumptions for both age groups when taking into account a unmeasured time-varying confounder, i.e. cholesterol levels. We further investigated the impact of possible positivity violations on the results. For this analysis, 61 subjects with limited overlap were discarded. The results did not differ significantly from the main analysis, although the effect was somewhat attenuated for both middle aged and older. 

%The posterior mean episodic memory score was attenuated both under the natural course and under the intervention when using the MNARS assumption instead of the MARS for the outcome. However, the estimates of the $PCIE$ were stable. 
%Finally, the results did not differ significantly from the main analysis when fitting the two other \textit{soft} BART models, i.e. S-DART and S-DART-lag1, instead of LDART. However, the effects were much smaller for both middle aged and older using SDART-lag1 and for SDART. 
\begin{table}[H]
\renewcommand\thetable{A2}
\centering
\caption{Results for the $PCIE$ from the sensitivity analyses of the Betula data using the proposed approach. MARS refers to the default settings as described in Section 7. \textit{Unmeasured confounding} refers to a sensitivity analysis for assumption A2, \textit{Positivity} refers to a sensitivity analysis for A3, and MNARS refers to a sensitivity analysis A4b. SDART refers to the standard SoftBART and SDART-lag1 referes to SDART incorporating only the predictors from the last time point. See Section 7.2 for further details.}
\begin{tabular}{lcc}
\hline
 & Middle aged & Older \\ 
\hline
  MARS & 0.6 (-0.4; 1.8) & 1.8 (-0.2; 4.0) \\ 
  \hline
\textit{Sensitivity analyses} & &  \\
\hline
  Unmeasured confounding & 0.6 (-0.5; 1.7) & 1.7 (-0.4; 3.9) \\ 
  Positivity & 0.4 (-0.5; 1.5) & 1.4 (-0.6; 3.4) \\ 
  MNARS & 0.6 (-0.5; 1.7) & 1.7 (-0.4; 3.9) \\ 
  SDART & 0.2 (-0.7; 1.3) & 0.7 (-1.4; 3.2) \\ 
  SDART-lag1 & 0.0 (-0.8; 1.1) & 0.1 (-1.2; 1.9) \\ 
  \hline
\end{tabular}
\label{table_betula_SA}
\end{table}

%\end{document}

\end{document}